\documentclass[9pt,twocolumn,twoside]{osajnl}
\usepackage{babel}
\journal{ol}
\setboolean{shortarticle}{true}

\title{Wiggling Instabilities of Temporal Localized States in Passively Mode-Locked Vertical External-Cavity Surface-Emitting Lasers}

\author[1,2,*]{Denis Hessel}
\author[1,2,3]{Svetlana V. Gurevich}
\author[2]{Julien Javaloyes}

\affil[1]{Departament de F\'{\i}sica, Universitat de les Illes Balears \& IAC-3, Cra.\,\,de
Valldemossa, km 7.5, E-07122 Palma de Mallorca, Spain}
\affil[2]{Institute for Theoretical Physics, University of M\"unster, Wilhelm-Klemm-Str. 9, 48149 M\"unster, Germany}
\affil[3]{Center for Nonlinear Science (CeNoS), University of M\"unster, Corrensstraße 2, 48149 M\"unster, Germany}

\affil[*]{Corresponding author: d.hessel@wwu.de}

\ociscodes{\textcolor{red}{140.4050,190.5530,190.4370}}

\graphicspath{{./}{./figures/}} 

\begin{abstract}
We analyze the emergence of wiggling temporal localized states in a passively mode-locked Vertical External-Cavity Surface-Emitting Laser composed by a gain chip and a resonant saturable absorber mirror. We show that the wiggling instability stems from the interplay between the third-order dispersion induced by the micro-cavities and their frequency mismatch. The latter is identified as an experimentally crucial parameter defining the range of existence of stable emission. We reveal the homoclinic scenario underlying the wiggling phenomenon and we show how it allows to control the oscillation.
\end{abstract}

\setboolean{displaycopyright}{true}

\begin{document}

\maketitle


Among the wealth of dynamical behaviour encountered in photonic systems, wiggling (or zigzagging) motion of optical pulses has been widely observed both experimentally and theoretically in systems in which external influences like e.g., guiding, trapping or feedback loops are introduced, or when interactions such as those found in soliton molecules lead to dynamics.

In this Letter we investigate the spontaneous formation and the dynamics of wiggling temporal localized states (TLSs) in a \emph{time invariant} passively mode-locked (PML) system based upon a Vertical External-Cavity Surface-Emitting Laser (VECSEL) whose long cavity \cite{MJB-PRL-14,MJC-JSTQE-16,CJM-PRA-16,CSV-OL-18} is closed by a resonant saturable absorber mirror. The understanding of this wiggling instability is of paramount importance to inform upon the proper parameter ranges in which stable TLSs exist, thereby leading to a minimum time and amplitude jitter and better defined frequency combs.

We show that the interplay between the third order dispersion (TOD) induced by the micro-cavities, namely the \textonehalf{}-VCSEL gain mirror and the saturable absorber mirror, as well as their relative detuning, is essential for the onset of the wiggling phenomenon that was observed in~\cite{SCP-PRL-19}. Since the frequency mismatch between cavities is mostly set by design and can only be marginally modified thermally, we stress its design value to be of critical importance. For both positive and negative detuning values we reveal the existence of a homoclinic bifurcation of limit cycles allowing to manipulate both the period and the amplitude of the wiggling. In addition, we identify that the detuning controls the appearance of additional Andronov-Hopf instabilities that further limit the regimes of existence of stable TLSs.
We employ a first principle model~\cite{MB-JQE-05,MJB-JSTQE-15,SJG-OL-18,SCP-PRL-19} based on delay algebraic equations (DAEs). The latter have shown recently their relevance in the modelling of dispersive phenomena in optical systems~\cite{SCP-PRL-19}. The proper consideration of coupled, dispersive, micro-cavities is essential to the scenario described in this work.

More generally, wiggling behaviour was observed during propagation in optical potentials generated by photovoltaic photorefractive crystals~\cite{ZCLLX-JOSAB-16} and in waveguides with either a triangular~\cite{SvG-JNLOP-01} or a parabolic index profile~\cite{DW-APB-06}. Zigzagging dissipative solitons were also found in a wide-aperture laser with a saturable Absorber (SA) subjected to optical feedback~\cite{PVGY-PRA-16} or during the fission of higher-order solitons impinging upon a lattice potential~\cite{KCZVSLT-PRL-04}. Finally, a secondary instability of the so-called creeping solitons observed in the cubic-quintic complex Ginzburg-Landau equation~\cite{SCAA-PRL-00,ASCT-PRE-01,CAASC-PRE-07} also leads to an apparent zigzag motion.
Wiggling dynamics may also stem from interacting localized states such as, e.g., the soliton molecules observed in a PML erbium-doped fiber laser~\cite{KNATDG-PRL-17} or the bound states of light bullets found in diffractive PML lasers~\cite{DJG-Chaos-20}. The \emph{intrinsic} wiggling of dissipative soliton was only recently observed experimentally by employing time-lens and dispersive Fourier transform techniques~\cite{ZCHTL-OL-20} in a fiber PML laser. Beyond photonics, the wiggling of localized states in chemical and soft-matter systems was also reported in, e.g.,~\cite{SOMS-PRE-95,OKGT-Chaos-20}.

\begin{figure}[t]
\center
\includegraphics{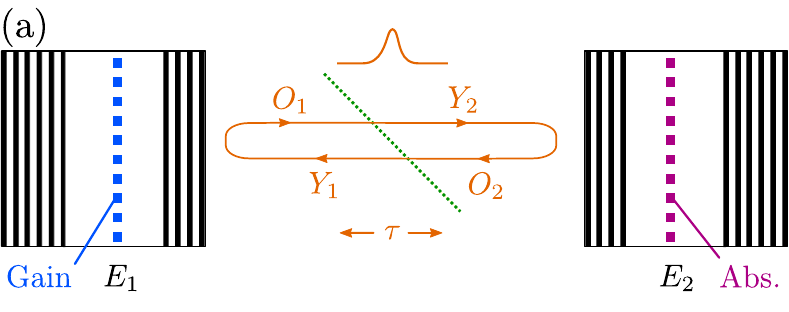}
\hspace{-0cm}\includegraphics{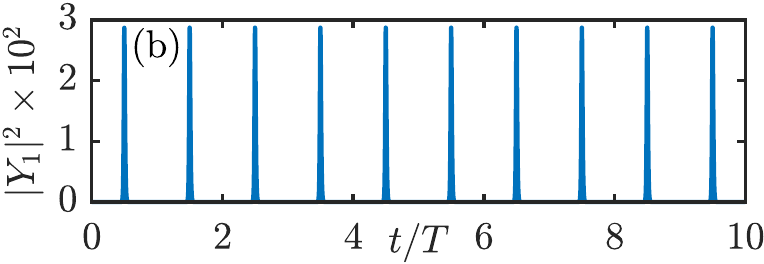}
\caption{(a) A schematic of the coupled cavities configuration. $E_i$ denote the intracavity fields, $i=1,\,2.$. The input and output fields in the external cavity are represented by $Y_i$ and $O_i$, respectively, whereas $\tau$ is the time of flight in the cavity. (b) A typical time trace found by integrating Eqs.~(\ref{eq:vecsel_e1})-(\ref{eq:vecsel_y2}) for  $j = 0.5406$; in the long cavity limit, stable pulsing in the form of TLSs occurs below the CW threshold and $T\simeq 2\tau$.}\label{fig:setup_and_traces}
\end{figure}



\begin{figure}[t]
\center
\includegraphics[width=0.5\textwidth]{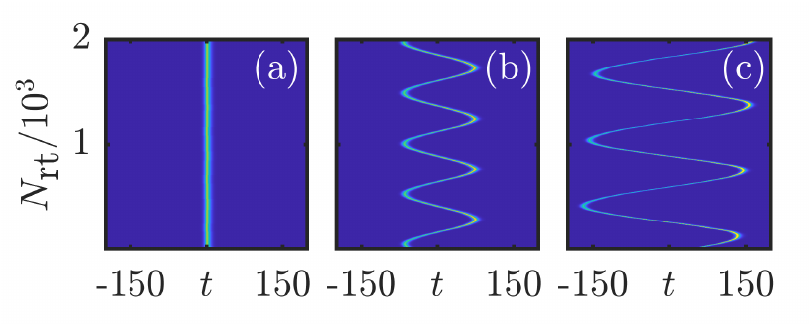}
\caption{Pseudo space-time diagrams for the single pulse train found by direct numerical simulations of Eqs.~(\ref{eq:vecsel_e1})-(\ref{eq:vecsel_y2}). The  intensities $|Y_1|^2$ are shown. (a) Stable TLS; (b) Wiggling TLS; (c) The period and amplitude of the TLS oscillation grow arbitrarily large. Parameters are $\delta = -0.5$ and $j =(0.5406,~0.5398,~0.5356)$ for (a), (b) and (c) respectively.}\label{fig:space_times}
\end{figure}

The schematic setup of our VECSEL system is depicted in Fig.~\ref{fig:setup_and_traces}~(a). It consists of two micro-cavities: a \textonehalf{}-VCSEL gain mirror and a resonant SA mirror separated by a time of flight $\tau$. The fields in the micro-cavities are denoted $E_i$, where $i=1,\,2$ correspond to the gain and the absorber mirrors, respectively. The output field $O_i$ of each nonlinear mirror turns into the injection field $Y_i$ of the other after passing a beam-splitter that is used to extract a signal.  The system is operated in the regime where the round-trip time is much longer than the semiconductor gain recovery time. 
In this so-called long cavity regime, TLSs can be observed below the continuous wave (CW) threshold~\cite{MJB-PRL-14,SJG-OL-18}. One typical time trace corresponding to stable fundamental mode-locking is depicted in Fig.~\ref{fig:setup_and_traces}~(b). There, the shape of the pulse does not change from round-trip to round-trip as can be seen more easily in the pseudo space-time representation in Fig.\ref{fig:space_times}~(a).

Following the approach developed in~\cite{MB-JQE-05,MJB-JSTQE-15,SJG-OL-18,SCP-PRL-19}, one can write the dynamical model for the intracavity fields $E_i$ and population inversions $N_i$ as 
\begin{align}
\kappa_{1}^{-1}\dot{E}_1 &= \left[(1-i\alpha_1)N_1-1\right]E_1 + h_1 Y_1\,, \label{eq:vecsel_e1}\\
\kappa_{2}^{-1}\dot{E}_2 &= \left[(1-i\alpha_2)N_2-1+i\delta\right]E_2 + h_2 Y_2\,, \\
\dot{N}_1 &= \gamma_1 (J_1 - N_1) - |E_1|^2 N_1\,, \\
\dot{N}_2 &= \gamma_2 (J_2 - N_2) - s|E_2|^2 N_2\,. 
\end{align}
Here, $\kappa_i^{-1}$ are the photon lifetimes, $\alpha_i$ are the linewidth enhancement factors and $\delta$ is the detuning between the two micro-cavities, $J_i$ are the bias in the gain and the absorber sections and $\gamma_i$ are the corresponding lifetimes. The ratio of the gain and absorber saturation intensities is $s$. The field injected into a micro-cavity is denoted with $Y_i$ whilst the coupling parameters $h_i\in[0,2]$ depend on the cavity mirror reflectivites.
The two micro-cavities mutually inject each other and their outputs consist in a superposition between the reflected and emitted fields. The link between the two micro-cavities, considering all multiple reflections, is given by two DAEs that physically correspond to the boundary conditions linking the fields defined in the three cavities composing the system,
\begin{align}
Y_1(t) &= \eta \left[E_2(t-\tau) - Y_2(t-\tau)\right] \label{eq:vecsel_y1}\,,\\
Y_2(t) &= \eta \left[E_1(t-\tau) - Y_1(t-\tau)\right]\label{eq:vecsel_y2}\,, 
\end{align}
where $\eta$ is the amplitude transmission of the beam-splitter and the minus sign before the injected field represents a phase shift of $\pi$ upon reflection from the top Bragg mirror.

In the following, we operate in the regime of localization, where the pulses are TLSs that appear below the lasing threshold bias. We denote with $j =  J_1/J^{\textrm{max}}_{\textrm{th}}$ the gain bias scaled with the maximum value of the threshold for CW emission, i.e. $J^{\textrm{max}}_{\textrm{th}} = \max J_{\textrm{th}}(\delta)$, where $J_{\textrm{th}}(\delta)$ is the CW threshold depending on $\delta$. The CW threshold is obtained by making the ansatz $E_j = E_{j,0} \exp(-i\omega t)$ and $Y_j = Y_{j,0} \exp(-i\omega t)$  which relates the output $O_{1,2} = r_{2,1} Y_{2,1}$ of one cavity to the injected field into the other cavity. In the long delay limit the threshold is computed by imposing the round-trip reflectivity of the full system to be unity, i.e. $R(\omega,J_1) = |r_1| |r_2| \eta^2$, with $r_i$ the unsaturated coefficients of reflection of the gain and absorber mirrors, respectively. We find that $r_1 = h_1/(1-i\omega/\kappa_1-J_1(1 -i\alpha_1))-1$, $r_2 = h_2 /(1-i (\delta+\omega/\kappa_2)-J_2(1-i\alpha_2))-1$. In the long cavity regime the lasing occurs very close to the frequency $\omega_{\textrm{max}}$ that maximizes  $R(\omega,J_1)$.

The parameters of both micro-cavities are set to $(\kappa_1, \alpha_1, \gamma_1) = (1, 2.5, 4.3\times 10^{-4})$ and $(\kappa_2, \alpha_2, J_2, \gamma_2) = (4.2857, 1.0, -0.07, 6.9\times 10^{-3})$, respectively~\cite{SJG-OL-18}. We choose the time of flight $\tau=5000$, where $2\tau$  corresponds to one round-trip. The other parameters are $(s, \eta) = (5, 0.99)$. The gain bottom Bragg mirror was assumed to be perfectly reflective giving $h_1 = 2$ while for the absorber, $h_2 = 1.9985$ was used to model the presence of non-saturable losses. We remark that $h_1=2$ corresponds to the case of an ideal Gires-Tournois interferometer~\cite{GT-CRA-64}. The latter are designed to conserve the photon number using highly reflective bottom mirrors and therefore yield a  purely dispersive spectrum \cite{SCP-PRL-19}. The models  based upon DAEs uch  as  those  given  by  Eqs.~(\ref{eq:vecsel_e1})-(\ref{eq:vecsel_y2}) correctly reproduce this unitary, dispersive, response \cite{SCP-PRL-19}.

The long cavity and the multiscale nature of the problem render the direct numerical simulations particularly tedious. To circumvent this difficulty, we use the functional mapping method~\cite{SJG-OL-18}. With this approach Eqs.~(\ref{eq:vecsel_e1})-(\ref{eq:vecsel_y2}) are integrated only in the vicinity of the so-called \emph{fast stage} of the pulse where the intensity is large and stimulated emission and absorption are the dominant effects. For the interpulses segments of the periodic solution, the so-called \emph{slow stage} where $E_j \approx 0$, the carrier populations recover exponentially, which can be readily solved analytically. This allows skipping the integration of the much longer slow stage. Finally, the subsequent slow and fast stages are simply connected by boundary conditions that reflect the continuity of the solution, see \cite{SJG-OL-18} for more details. For all direct numerical simulations the length of the integration time box in the vicinity of the pulse was set to $t_{\textrm{box}} = 400$.
The functional mapping method allows us to perform efficient parameter scans for the single temporal LS regime and we present some of our results in Fig.~\ref{fig:space_times}. We start with the parameter set yielding a stable pulsating regime (cf. Fig.~\ref{fig:setup_and_traces}~(b)) and its pseudo space-time representation is presented in Fig.~\ref{fig:space_times}~(a). There, we used as a folding parameter the exact period of the solution which results in a vertical spatio-temporal trace. As the scaled gain bias $j$ is decreased, keeping all other parameters fixed, the pulse starts to oscillate in its amplitude and position, see Fig.\ref{fig:space_times}~(b). We refer to these oscillations where the pulse moves back and forth without a net drift as \emph{wiggling}. Remarkably, by further decreasing $j$, the period and the amplitude of the wiggling oscillations can be made much larger letting $j$ approach a specific value $j_{\textrm{Hom}}$. Below $j_{\textrm{Hom}}$ only the off solution is stable. The temporal evolution of a TLS with $j$ even closer to $j_{\textrm{Hom}}$ can be seen in Fig.~\ref{fig:space_times}~(c) where the period and the amplitude of the  wiggling is remarkably larger than in panel (b), see also Visualisation 1.

\begin{figure}[t]
\center
\includegraphics[width=0.45\textwidth]{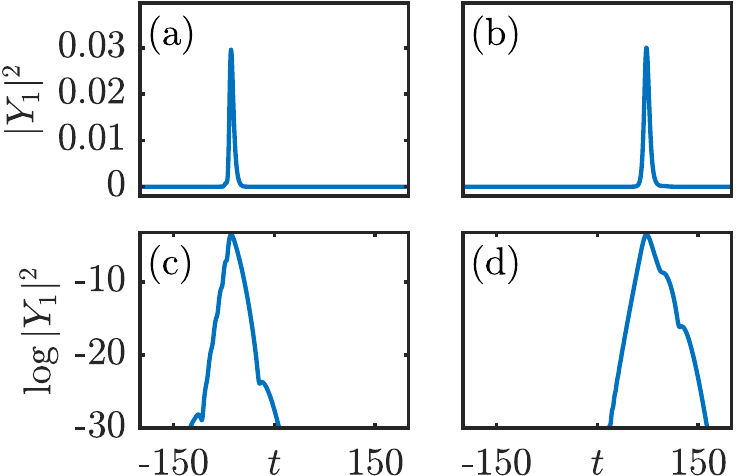}
\caption{Details on the pulse profile in the wiggling regime (linear and logarithmic scales) for the same parameters as in Fig.~\ref{fig:space_times}(c). Panels (a,c) and (b,d) correspond to the round-trip numbers 47 and 281 respectively.}\label{fig:trace_detail}
\end{figure}

It was shown in ~\cite{SCP-PRL-19}, that the third order dispersion stemming from the lasing micro-cavity induces a train of decaying satellites on the leading edge of the TLSs observed in this system. Due to the nonlinear interaction with carriers, these satellites may get amplified, eventually replacing the parent pulse that would die out. This regime was termed a satellite instability, see Fig.~4(a) in~\cite{SCP-PRL-19}. A similar phenomenon also appear in mode-locked integrated external-cavity surface emitting lasers~\cite{SHJ-PRA-20}, see Fig.~8. There, the gain and the absorber share the same microcavity, which leads to a somewhat simpler scenario.
In the present case, we note that the wiggling oscillations are not a satellite instability since the parent pulse remains fully merged with its leading satellite, see Fig.~\ref{fig:trace_detail}~(a,b), the latter could merely be observed using a logarithmic scale as shown in Fig.~\ref{fig:trace_detail}~(c,d). However, since this wiggling instability was also observed in ~\cite{SCP-PRL-19} Fig.~4(b), for slightly different parameters than for the satellite instability, we shall conclude that, in all cases, third order dispersion remains at the root of the observed oscillatory motion. Here, the emerging satellites immediately melt within the main pulse which creates an overall, apparent, wiggling motion.

In order to shed further light on the mechanism responsible for the wiggling, we performed a bifurcation analysis of ~(\ref{eq:vecsel_e1})-(\ref{eq:vecsel_y2}) using a recent, modified version of the continuation tool DDE-Biftool~\cite{DDEBT} adapted to DAEs.
\begin{figure}[t]
\center
\includegraphics{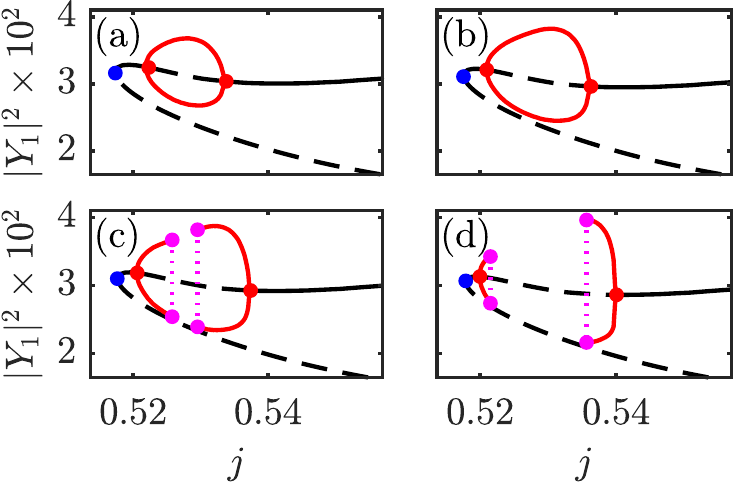}
\caption{The branches of single TLS solution, showing the peak intensity $|Y_1|^2$ as a function of the normalized gain $j_1$ for different values of $\delta = (-0.475, -0.485, -0.49, - 0.5)$ (a,b,c,d) superposing results from path continuation and direct numerical simulations. Solid (dashed) black line corresponds to the stable (unstable) TLS branch. Blue points mark the corresponding folds. The quasi-periodic branch that corresponds to wiggling TLSs is depicted in red. As the amplitude of the wiggling grows changing $\delta$, the inflating red branch breaks in two parts (c,d) each terminated by a homoclinic bifurcation.}\label{fig:fold_of_homoclinics}
\end{figure}
%
Figure~\ref{fig:fold_of_homoclinics}~(a) shows the part of a branch of a single TLS, where a maximum of the injected field intensity $|Y_1|^2$ as a function of the scaled gain bias $j$ for the fixed detuning $\delta$ slightly smaller than in Fig.~\ref{fig:space_times}.
The TLSs are periodic orbits of the DAEs~(\ref{eq:vecsel_e1})-(\ref{eq:vecsel_y2}) and they emerge in a saddle-node of limit cycles bifurcation at $j=j_F$ (blue circle).
The TLSs are unstable over the low power branch (dashed black line) and stable on the upper branch close to the fold if $j$ is increased (solid black line). 
However, for increasing $j$, an Andronov-Hopf bifurcation appears at $j=j_T$ (red circle) and a quasi-periodic solution, that corresponds to a wiggling TLS, emerges. 
Since the path-continuation of quasiperiodic orbits is not possible within DDE-Biftool, we conducted direct numerical simulations to reconstruct this orbit (red line). 
Here, the maximum and minimum intensity values per round-trip as represented. One can clearly see that the branch corresponding to wiggling TLSs emerges supercritically and connects two torus bifurcation points. Further increasing $j$, the high power branch of TLS recovers its stability until $j=j_{\mathrm{th}}$.
If the magnitude of $|\delta|$ is increased, the wiggling orbit grows, see Fig.\ref{fig:fold_of_homoclinics}~(b), until its lower part touches the unstable TLS branch. 
This point $j=j_{\textrm{Hom}}$ corresponds to a fold of two emerging \emph{homoclinic bifurcations of periodic orbits}. 
If $|\delta|$ is further increased beyond this point (see Fig.\ref{fig:fold_of_homoclinics}~(c)), the torus orbit splits into two parts, each limited by a homoclinic bifurcation point (magenta circles) and a homoclinic orbit connecting these points, which is schematically depicted in dashed magenta line. Note that in the interval between the two dashed magenta lines, no stable solution exists and the system converges to the off state. For larger values of $|\delta|$, the torus bifurcation points slowly move away from each other along the TLS branch, whereas the homoclinic points are moving towards the $j_T$ points, as presented in Fig.\ref{fig:fold_of_homoclinics}~(d). Now the behaviour observed in Fig.~\ref{fig:space_times} becomes feasible: For high gain bias values, a TLS is stable, see Fig.\ref{fig:fold_of_homoclinics}~(d) and Fig.~\ref{fig:space_times}~(a). Decreasing $j$, a torus bifurcation sets in at $j=j_T$ and a wiggling TLS occurs (cf. Fig.~\ref{fig:space_times}~(b)). Further decreasing $j$ one moves along the torus orbit and the oscillation period grows (Fig.~\ref{fig:space_times}~(c)) and can be arbitrarily large approaching the homoclinic point $j=j_{\textrm{Hom}}$, where the period becomes theoretically infinite. 
\begin{figure}[t]
\center
\includegraphics{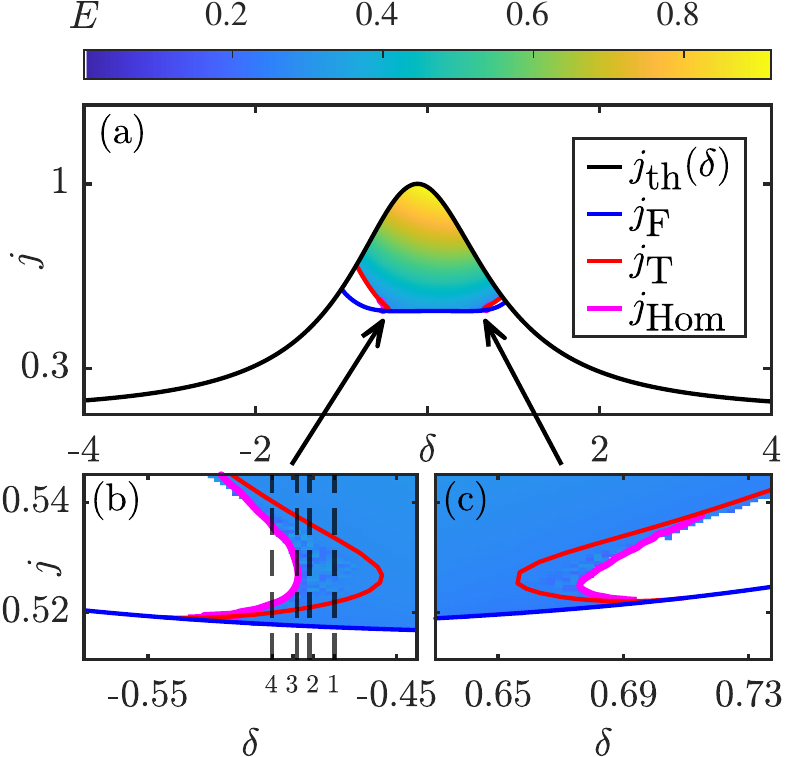}
\caption{(a) Two-parameter bifurcation diagram in the $(j,\,\delta)$ plane together with the pulse energy $E$ of the injected field $Y_1$ obtained from direct numerical simulations of Eqs.~(\ref{eq:vecsel_e1})-(\ref{eq:vecsel_y2}). (b) and (c): Insets in the vicinity of the left and right co-dimension two point, respectively. Black, blue, red and magenta solid lines indicate  the threshold for continuous wave emission $j_{\mathrm{th}}$, the fold of periodic orbits $j_F$, torus bifurcation line as well as homoclinic bifurcation line of periodic orbits, respectively. Black dashed lines from right to left correspond to the cross-sections depicted in Figs.~\ref{fig:fold_of_homoclinics}~(a)-(d), respectively.   
}\label{fig:two_parameter_bif}
\end{figure}


Finally, in Fig.~\ref{fig:two_parameter_bif} we present a two-parameter study that reveals the interplay between the value of the bias current and the detuning between the two micro-cavities. We depict the pulse energy $E$ of the $Y_1$ field obtained from direct numerical simulations of Eqs.~(\ref{eq:vecsel_e1})-(\ref{eq:vecsel_y2}) in the $(j,\,\delta)$ plane. Intuitively, one understands that the region of existence of the TLS solutions must be bounded between the threshold line $j_{\mathrm{th}}$ (black solid line)-- where the (background) off solution becomes unstable -- and the fold of period orbits $j_{F}$ (blue solid line) where the TLS branch emerges.
However, studying the stability of the TLS solution imposes more stringent conditions and reveal the importance of the detuning $\delta$. While for small $|\delta|$ the TLS is stable for all the bias values between the two aforementioned bordering lines, two unstable regions for both positive and negative detunings appear if $|\delta|$ is increased. There, the onset of stable TLS emission is governed by the torus (red) lines and the homoclinic bifurcations (magenta). Both lines collide at one point in the $(j,\,\delta)$  plane where they meet with the fold line. This point corresponds to a so-called co-dimension two point, which is the analogon of a Bogdanov-Takens bifurcation point found in dynamical systems governed by ordinary differential equations. The branches presented in Fig.~\ref{fig:fold_of_homoclinics} can be seen as four cross-sections of the Fig.~\ref{fig:two_parameter_bif}~(b,c) for different $\delta$ values, that are indicated by the four dashed black labeled lines; the labels (from right to left) correspond to the panels (a)-(d) of the Fig.~\ref{fig:fold_of_homoclinics},


In conclusion, we analyzed in this manuscript how wiggling temporal localized states can spontaneously appear in the framework of a passively mode-locked Vertical External-Cavity Surface-Emitting Lasers (VECSELs). The wiggling results from the interplay between the third order dispersion stemming from the micro-cavity and their respective detuning. The latter is identified as an experimentally crucial design parameter that can defines the range of existence of stable TLSs. The wiggling is the results of the interaction between the pulse and its emerging unstable satellites with which the pulse further coalesces thereby creating an apparent motion that fully explains the results obtained by \cite{SCP-PRL-19}. Further, we revealed the existence of a homoclinic bifurcation of limit cycles allowing for a controllable tuning of the wiggling oscillation period.


\section*{funding}
J.J. acknowledges the financial support of the MINECO Project MOVELIGHT (PGC2018-099637-B-100 AEI/FEDER UE). 
\section*{Disclosures}
The authors declare no conflicts of interest


\end{document}